\documentclass[lettersize,journal]{IEEEtran}
\usepackage{amsmath,amsfonts}
\usepackage{algorithm}
\usepackage{array}
\usepackage[caption=false,font=normalsize,labelfont=sf,textfont=sf]{subfig}
\usepackage{textcomp}
\usepackage{stfloats}
\usepackage{url}
\usepackage{verbatim}
\usepackage{graphicx}
\usepackage{cite}
\usepackage{algorithmicx}
\usepackage[noend]{algpseudocode}

\newcommand{\DefineClass}[1]{\item[\textbf{Class:}] #1}

\usepackage{xcolor}

\makeatletter
\def\ps@IEEEtitlepagestyle{%
  \def\@oddhead{%
    \parbox{\textwidth}{\centering\footnotesize
\color{gray} \copyright 2025 IEEE.  Personal use of this material is permitted.  Permission from IEEE must be obtained for all other uses, in any current or future media, including reprinting/republishing this material for advertising or promotional purposes, creating new collective works, for resale or redistribution to servers or lists, or reuse of any copyrighted component of this work in other works. \color{black}
    }
  }
  \def\@evenhead{\@oddhead}%
  \def\@oddfoot{}%
  \def\@evenfoot{}%
}
\makeatother

\title{\LARGE \bf  
MAPF-HD: Multi-Agent Path Finding\\in High-Density Environments
}

\author{Hiroya Makino$^{1,*}$ and Seigo Ito$^{1}$
\thanks{$^{1}$ H. Makino and S. Ito are with Toyota Central R\&D Labs., Inc., 41-1, Yokomichi, Nagakute, Aichi, Japan.}
        \thanks{$^{*}$ Corresponding author. {\tt\small hirom@mosk.tytlabs.co.jp}}%
        }

\begin{document}

\maketitle
\pagestyle{empty}

\begin{abstract} 
Multi-agent path finding (MAPF) involves planning efficient paths for multiple agents to move simultaneously while avoiding collisions.
In typical warehouse environments, agents are often sparsely distributed along aisles; however, increasing the agent density can improve space efficiency.
When the agent density is high, it becomes necessary to optimize the paths not only for goal‑assigned agents but also for those obstructing them.
This study proposes a novel MAPF framework for high-density environments (MAPF-HD). 
Several studies have explored MAPF in similar settings using integer linear programming (ILP). 
However, ILP-based methods require substantial computation time to optimize all agent paths simultaneously. 
Even in small grid-based environments with fewer than $100$ cells, these computations can take tens to hundreds of seconds. 
Such high computational costs render these methods impractical for large-scale applications such as automated warehouses and valet parking. 
To address these limitations, we introduce the phased null-agent swapping (PHANS) method. 
PHANS employs a heuristic approach to incrementally swap positions between agents and empty vertices. 
This method solves the MAPF-HD problem within a few seconds, even in large environments containing more than $700$ cells. 
The proposed method has the potential to improve efficiency in various real-world applications such as warehouse logistics, traffic management, and crowd control.
\end{abstract}

\section{Introduction}
\IEEEPARstart{I}{ndustry} is increasingly adopting automated guided vehicles (AGVs) and conveyors to enhance efficiency and reduce reliance on human labor.
In environments with agents, such as AGVs or pallets on conveyors, efficient collision-free path planning is essential.
These path-planning problems for multiple agents are formalized as multi-agent path finding (MAPF) problems~\cite{stern2019}.
MAPF has been applied to various settings, including automated warehouses~\cite{wurman2008}, airport taxiway management~\cite{li2019}, automated parking systems~\cite{okoso2019}, and video games~\cite{silver2005}.

\begin{figure}[!t]
  \centering
  \includegraphics[width=0.96\linewidth]{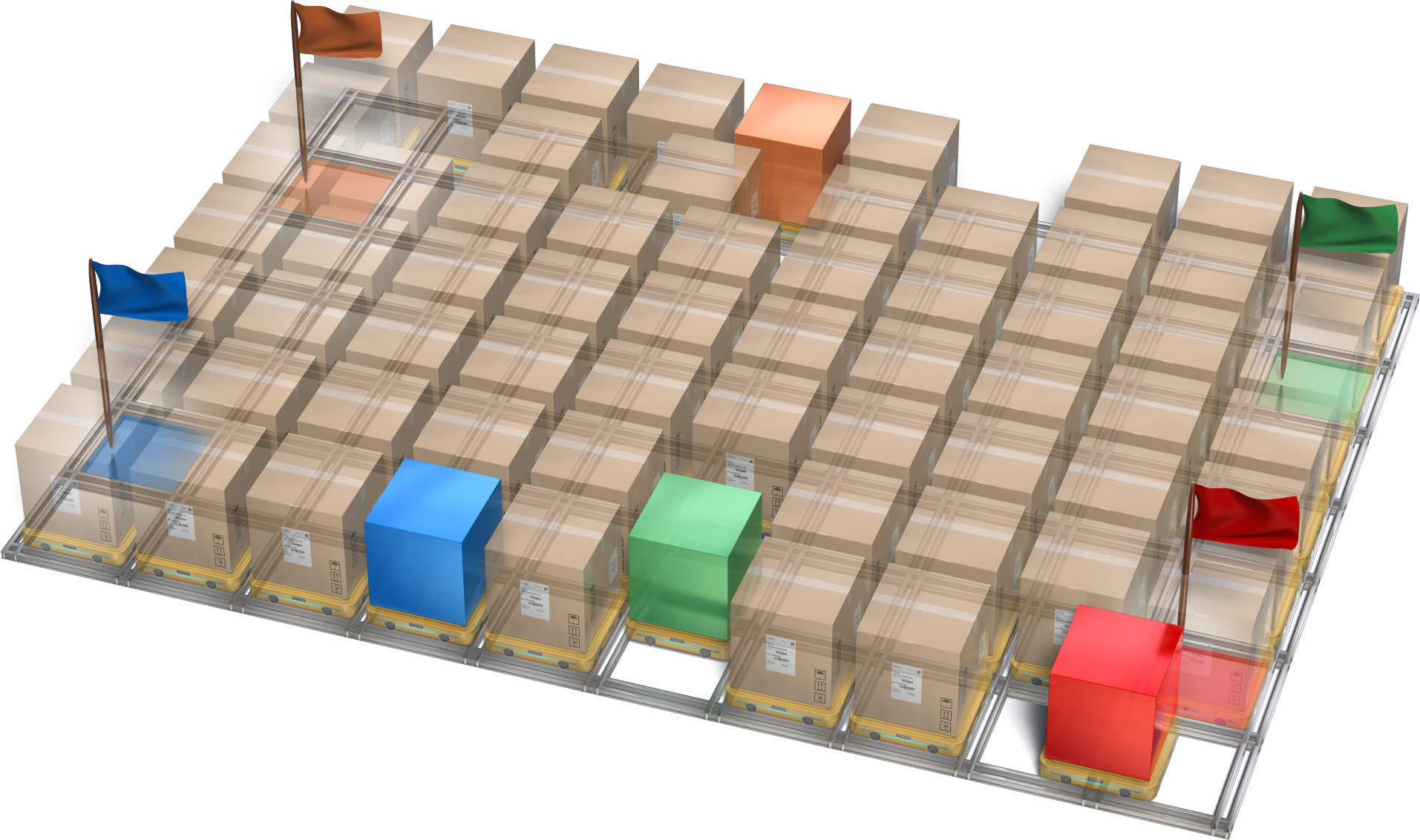}
  \caption{MAPF in a high-density warehouse environment. Only a subset of target agents (those with matching-colored flags) are assigned goals, whereas the rest act as movable obstacles. In addition to planning paths for target agents to reach their goals, efficiently relocating obstructing agents is crucial for successful navigation under high-density conditions.}
  \label{cover}
\end{figure}

In typical MAPF settings, agents are often sparsely distributed along aisles; however, increasing the agent density can improve space efficiency.
For instance, high-density automated warehouses store significantly more goods than lower-density ones, reducing required floor space and operational costs (Fig.~\ref{cover}).
Similarly, high-density automated valet parking systems allow vehicles to be parked and retrieved in a smaller area by minimizing the gaps between vehicles.

Despite these space-saving benefits, planning efficient paths using conventional techniques remains challenging.
Typically, in high-density environments, only a subset of agents has active goals at any given time (e.g., moving some items for shipment). 
The remaining agents, although not assigned to any particular destination, still occupy space and obstruct the movement of the target agents. 
For the target agents to reach their objectives, the obstructing agents must be relocated efficiently.
One approach is to use integer linear programming (ILP) to plan all agent paths jointly~\cite{okoso2022, makino2024}. 
However, this approach generally incurs significant computation time because all paths are optimized simultaneously by ILP.
Another approach is to set temporary goals for obstructing agents and then apply existing MAPF methods. 
However, this often extends the paths of the target agents because the relocation of the obstructing agents is not optimized. 
Table \ref{tab1} highlights these limitations and the need for a tailored solution to accommodate obstructing agents efficiently.

\begin{table}[!t]
    \centering
    \caption{Methods for MAPF in high-density environments.}
    \label{tab1}
    \footnotesize
    \begin{tabular}{|l|lll|}
    \hline
                & Target                                                & Obstructing                                           & Computational                             \\ 
    Methods     & agents                                                & agents                                                & efficiency                                \\ \hline
    ILP         & \color{green}$\checkmark$\color{black} (optimal)      & \color{green}$\checkmark$\color{black} (optimal)      & \color{red}$\times$\color{black}          \\
    Heuristics  & \color{green}$\checkmark$\color{black} (suboptimal)   & \color{red}$\times$\color{black} (not optimized)      & \color{green}$\checkmark$\color{black}    \\
    PHANS       & \color{green}$\checkmark$\color{black} (suboptimal)   & \color{green}$\checkmark$\color{black} (suboptimal)   & \color{green}$\checkmark$\color{black}    \\ \hline
    \end{tabular}
\end{table}

\begin{figure*}[!t]
  \centering
  \includegraphics[scale=0.12]{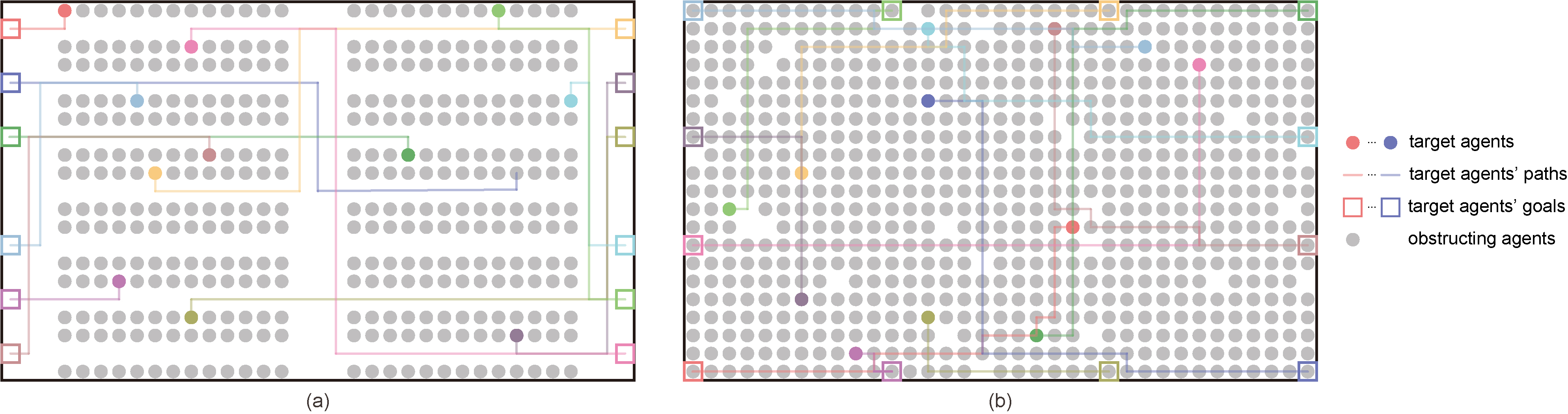}
  \caption{Simulation examples of MAPF in (a) low-density and (b) high-density environments. Colored lines represent the paths of the target agents. Unlike low-density environments, obstructing agents are present along these paths in high-density environments. The target agents are moved while the obstructing agents are appropriately evacuated.}
  \label{mapf_and_mapfh_env}
\end{figure*}

This study refines MAPF for high-density environments (MAPF-HD) and develops an efficient algorithm tailored to this context.
This algorithm employs phased heuristic techniques to relocate agents by swapping their positions with empty vertices (null agents).
This approach, termed phased null-agent swapping (PHANS), solves MAPF-HD within a few seconds even in a densely packed environment, as shown in Fig.~\ref{mapf_and_mapfh_env}.
The implementation is available at \color[HTML]{0000ff}\url{https://github.com/ToyotaCRDL/MAPF-in-High-Density-Envs}\color{black}.

\section{Related Work}
\subsection{Optimal Solvers}
MAPF involves determining optimal collision-free paths for multiple agents~\cite{stern2019, felner2017}.
Conflict-based search (CBS)~\cite{sharon2015}, a widely used optimal solver, plans agents' paths to their goals (low-level search) and resolves collisions by branching (high-level search). 
However, CBS faces scalability challenges because MAPF is NP-hard~\cite{yu2013a}.
Researchers have introduced improvements, such as Improved CBS, to prune the search more effectively~\cite{boyarski2015}.
Other optimal frameworks recast MAPF as a combinatorial optimization problem. 
For example, Lam et al.~\cite{lam2019} applied the branch-and-cut-and-price ILP method.
Similarly, G\'omez et al.~\cite{gomez2020} used answer-set programming.

\subsection{Suboptimal Solvers}
Several bounded-suboptimal and heuristic algorithms have been developed to improve the scalability of MAPF~\cite{gao2024}. 
Enhanced CBS (ECBS)~\cite{barer2014} is a bounded-suboptimal solver that employs focal search. 
To enhance scalability, explicit estimation CBS (EECBS)~\cite{li2021c} replaces the focal search in ECBS with explicit estimation search.
Priority inheritance with backtracking (PIBT)~\cite{okumura2022}, a decentralized MAPF algorithm, assigns dynamic movement priorities to agents and allows backtracking to resolve conflicts. 
Unlike static priority planning, PIBT enables agents to inherit and adjust priorities dynamically, preventing deadlocks and improving coordination in dense environments. 
Backtracking allows agents to reconsider past actions when conflicts arise, thereby ensuring flexible and adaptive navigation.

PIBT is both scalable and applicable to high-density environments.
Chen et al.~\cite{chen2024b} scaled PIBT for 2,000--12,000 agents occupying $90\%$ of the available cells.
Jiang et al.~\cite{jiang2024b} showed that PIBT can solve a $32\times32$ map with $97.7\%$ agent density.

More recently, several reinforcement learning-based MAPF methods have been proposed~\cite{alkazzi2024c}.
These approaches require time for training; however, once a policy is learned, agents can select actions quickly even in large-scale environments.

\subsection{MAPF in High-Density Environments with Obstructing Agents}
Even in dense environments, relocation problems can be solved when all agents have assigned destinations.
However, to move target agents efficiently, the relocation of obstructing agents must be optimized.

Okoso et al.~\cite{okoso2022} introduced the cooperative automated valet parking problem (CoAVP).
In CoAVP, agents are classified into three types: entering, departing, and relocating vehicles.
Departing vehicles are assigned destinations.
Entering and relocating vehicles can be anywhere in the parking lot, and their destinations are also optimized.
Their ILP-based method extends the planar graph along the time axis and optimizes the paths within the time-expanded network.
However, this approach presents two major challenges. 
First, ILP has high computational complexity and requires hundreds of seconds even in small environments (e.g., $15\times15$ grids). Consequently, this method is impractical in real-world applications.
Second, collision types specific to high-density environments have not been sufficiently investigated.
To save parking space, agents must be densely positioned.
In such environments, two adjacent agents must move in almost perfect synchrony, which may not be realistic. 
Following conflicts~\cite{stern2019} should be considered for high-density environments.

Makino et al.~\cite{makino2024} defined a multi-agent and multi-rack path finding problem, where racks were classified as either target or nontarget.
Agents convey target racks to their destinations while relocating nontarget racks.
Similar to~\cite{okoso2022}, their approach uses ILP, resulting in high time complexity.

This study redefines MAPF in high-density environments and develops a heuristic approach to mitigate the high time complexity associated with ILP.

\section{Multi-Agent Path Finding Problem in High-Density Environments} 
\begin{figure}[!t]
  \centering
  \includegraphics[scale=0.12]{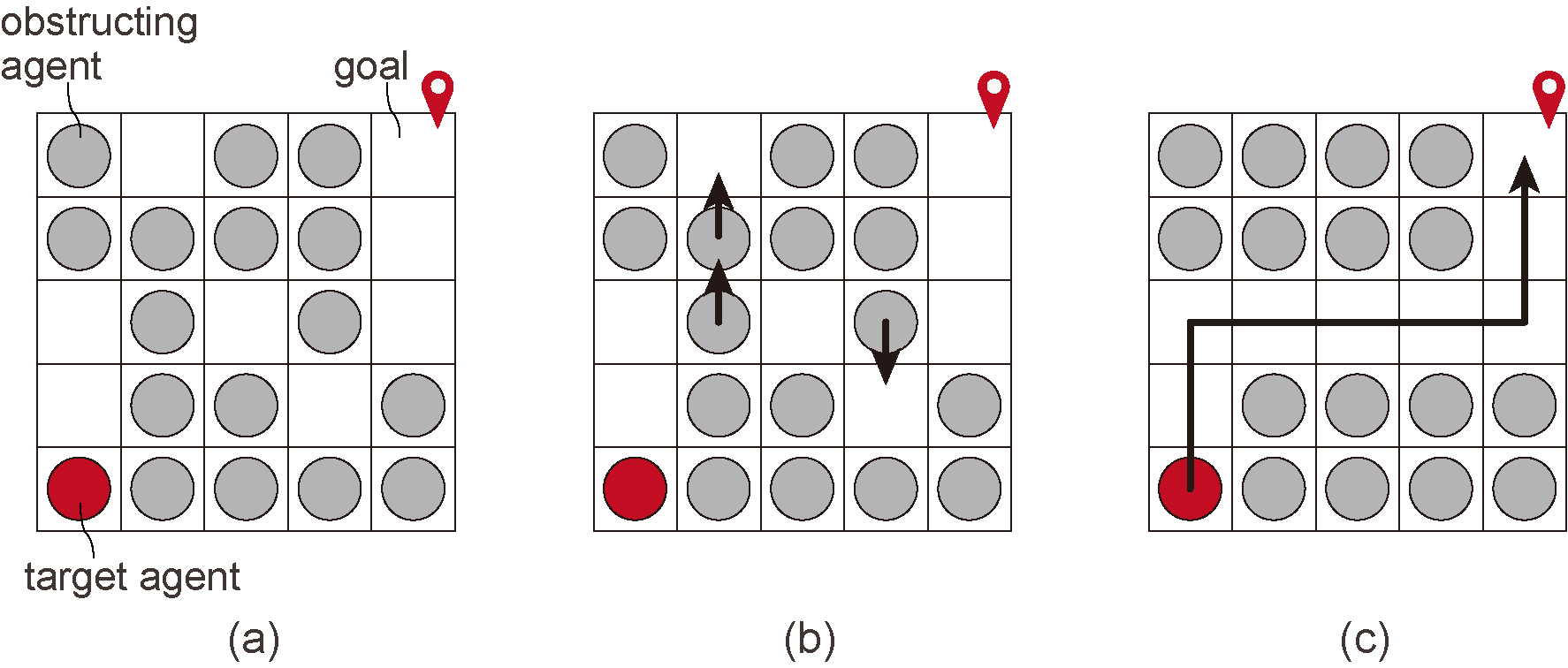}
  \caption{Example of the MAPF-HD problem: (a) initial positions, (b) evacuation of obstructing agents, and (c) movement of the target agent.}
  \label{mapfh_example}
\end{figure}

This section defines MAPF-HD. 
Unlike standard MAPF, which focuses on the paths of the target (goal-assigned) agents toward designated goals, MAPF-HD operates in environments where obstructing agents are densely packed.
The objective of MAPF-HD is to minimize the number of time steps required for all target agents to reach their goals without collisions.
Fig.~\ref{mapfh_example} shows an example of MAPF-HD.

\subsection{Environment} 
An instance of MAPF-HD includes a four-connected grid graph, $G=(V,E)$, where $V$ is the set of vertices, and $E$ is the set of edges. 
$V$ and $E$ are defined based on the grid dimensions $\mathrm{size}_x$ and $\mathrm{size}_y$ as follows:
\begin{align}
  &V=\left\{v_{x,y}\mid x \in \{0, \ldots, \mathrm{size}_x-1\}, y \in \{0, \ldots, \mathrm{size}_y-1\} \right\} \\
  &E=\left\{(v_{x,y}, v_{x^\prime,y^\prime})\mid v_{x,y}, v_{x^\prime,y^\prime} \in V, |x-x^\prime|+|y-y^\prime|=1 \right\}, 
\end{align}
where $x$ and $x^\prime$ are row indices, and $y$ and $y^\prime$ are column indices. 
These indices correspond to the grid coordinates.
The Manhattan distance $dst(v_{x,y}, v_{x^\prime,y^\prime})$ between two points is calculated as follows:
\begin{align}
  \label{eq:manhattan_distance}
  dst(v_{x,y}, v_{x^\prime,y^\prime}) = |x-x^\prime| + |y-y^\prime|.
\end{align}

\subsection{Agent} 
MAPF-HD involves two types of agents: 
\begin{itemize} 
\item Target agents: $A^\mathrm{tgt} = \{a^\mathrm{tgt}_1, \ldots, a^\mathrm{tgt}_m\}$, with $m > 0$. Each target agent $a^\mathrm{tgt}_i$ is specified by a tuple $\langle s^\mathrm{tgt}_i, g^\mathrm{tgt}_i \rangle$, where $s^\mathrm{tgt}_i \in V$ is the start vertex, and $g^\mathrm{tgt}_i \in V$ is the goal vertex. 
\item Obstructing agents: $A^\mathrm{obs} = \{a^\mathrm{obs}_1, \ldots, a^\mathrm{obs}_n\}$, with $n \geq 0$. Each obstructing agent $a^\mathrm{obs}_i$ is characterized solely by its start vertex, denoted as $\langle s^\mathrm{obs}_i \rangle$, as it does not have a goal. 
\end{itemize}
The initial positions of the target and obstructing agents do not overlap. 
Furthermore, the total number of agents is less than the number of vertices in the graph.

\subsection{Agent Action} 
Time is discretized into time steps, and at each time step $t$, every agent $a_i \in A = A^\mathrm{tgt} \cup A^\mathrm{obs}$ occupies a vertex $\pi_i[t] \in V$. 
At each time step, an agent executes one of the following actions: 
\begin{itemize}
  \item Remain at the current vertex: $\pi_i[t]=\pi_i[t+1]$.
  \item Move to an adjacent vertex: $\left(\pi_i[t], \pi_i[t+1] \right) \in E$.
\end{itemize}

Agents avoid collisions at each time step by managing the two conflicts:
\begin{itemize}
  \item Vertex conflict: Multiple agents simultaneously occupy the same vertex.
  \item Following conflict: An agent moves to a vertex occupied by another agent in the immediately preceding time step.
\end{itemize}
Some studies~\cite{sharon2015, li2021c, okumura2022} are based on edge conflicts~\cite{stern2019} rather than following conflicts. 
If following conflicts are not treated as conflicts, the agents must be spaced apart, and adjacent agents would need to move almost synchronously. 
However, in agent-dense environments, spacing the agents sufficiently apart may not be feasible. Moreover, perfect synchronization is not always realistic in real environments.
Therefore, we treat following conflicts as prohibited. 
Note that following conflicts contain edge conflicts~\cite{stern2019}.

To prevent these conflicts, all agents $a_i, a_j (i\neq j)$ must satisfy:
\begin{align}
  \begin{cases}
    \pi_i[t] \neq \pi_j[t] \\
    \pi_i[t] \neq \pi_j[t+1]. \\
  \end{cases}
\end{align}

\subsection{Solution and Objective} 
A solution to MAPF-HD is a set of collision-free paths $\{\pi_1, \ldots, \pi_{m+n} \}$ ensuring that all target agents reach their goals at a certain time step $T$.
None of the obstructing agents have goals, and they are freely positioned.
Formally, the problem assigns a path $\pi_i = [\pi_i[0], \pi_i[1], \ldots, \pi_i[T]]$ to each agent, which must satisfy the following:
\begin{align}
  \begin{cases}
      \pi_i[0] = s^\mathrm{tgt}_i \quad &\mathrm{if} \, \, a^\mathrm{tgt}_i \in A^\mathrm{tgt} \\
      \pi_i[0] = s^\mathrm{obs}_i \quad &\mathrm{if} \, \, a^\mathrm{obs}_i \in A^\mathrm{obs} \\
      \pi_i[T] = g^\mathrm{tgt}_i \quad &\mathrm{if} \, \, a^\mathrm{tgt}_i \in A^\mathrm{tgt}.
  \end{cases}
\end{align}

The objective is to minimize makespan $T$, which is defined as the total number of time steps required for all target agents to reach their goals.
In addition to makespan, classical MAPF often uses sum-of-costs~\cite{stern2019}, which is the sum of time steps required by each agent to reach its goal.
In real-world applications such as warehouses and valet parking, the emphasis is on the speed of the target agents in reaching their goals, rather than on the total movement effort of all agents. Makespan captures this operational bottleneck, whereas sum-of-costs would be distorted by the large number of obstructing agents without goals.

\section{Proposed Method}
Current methods~\cite{okoso2022, makino2024} use ILP to optimize the paths of all agents simultaneously, resulting in high computational complexity. 
To reduce the computational costs, we propose a two-stage heuristic approach termed PHANS.
For the target agents, the first stage plans paths from their starting positions to their goals using the A* algorithm.
The second stage sequentially evacuates the obstructing agents from these paths to clear the way for target agents.
Section IV-A describes the methods used in both stages through an example with a single target agent, followed by a generalized algorithm for multiple target agents in Section IV-B.

\subsection{Single Target Agent}
\subsubsection{First Stage}
The A* algorithm is used to plan a target path from the start to the goal vertices for the target agent.
Collisions with obstructing agents are allowed during this stage.
However, to minimize the costs of evacuating obstructing agents, the target agent's movement should satisfy the following conditions:
\begin{itemize}
\item Minimize the movement time of the target agent to ensure that the target path is as short as possible.
\item Minimize the waiting time for evacuating obstructing agents from the target path.
\end{itemize}
To satisfy the second condition, the A* algorithm is modified by introducing an additional heuristic value, $h_\mathrm{add}$.
This modification incorporates the expected waiting cost for obstructing agents to clear the path.
The cost function $f$ is computed as the sum of $g, h$, and $h_\mathrm{add}$ as follows:
\begin{align}
f = g + h + h_\mathrm{add},
\end{align}
where $g$ is the actual cost from the start to the current node, and $h$ is the estimated cost from the current node to the goal.
$h_\mathrm{add}$ is calculated as follows:
\begin{align}
h_\mathrm{add} = \max(0, 1 + h_\mathrm{evac} - g).
\end{align}
Here, $h_\mathrm{evac}$ estimates the cost of evacuating an obstructing agent by calculating its distance to the nearest empty vertex using \eqref{eq:manhattan_distance}.
If the estimate is accurate and the obstructing agents are evacuated by the time the target agent arrives, $h_\mathrm{add}$ is $0$.
Otherwise, the waiting time is incorporated into $h_\mathrm{add}$.
These computations are based on the initial positions; however, the set of empty vertices may change as obstructing agents are evacuated.

The proposed heuristic $h+h_\mathrm{add}$ is neither admissible nor consistent in general. 
When $h_\mathrm{add}=0$, it reduces to the standard admissible distance estimate.
In other cases, it may overestimate the true evacuation cost, and its value can change nonmonotonically between neighboring states. 
Thus, the heuristic serves as a practical cost estimator that reflects the additional effort of relocating obstructing agents in high‑density environments, rather than a guarantee of optimality.

\subsubsection{Second Stage}
\begin{figure}[!t]
  \centering
  \includegraphics[scale=0.12]{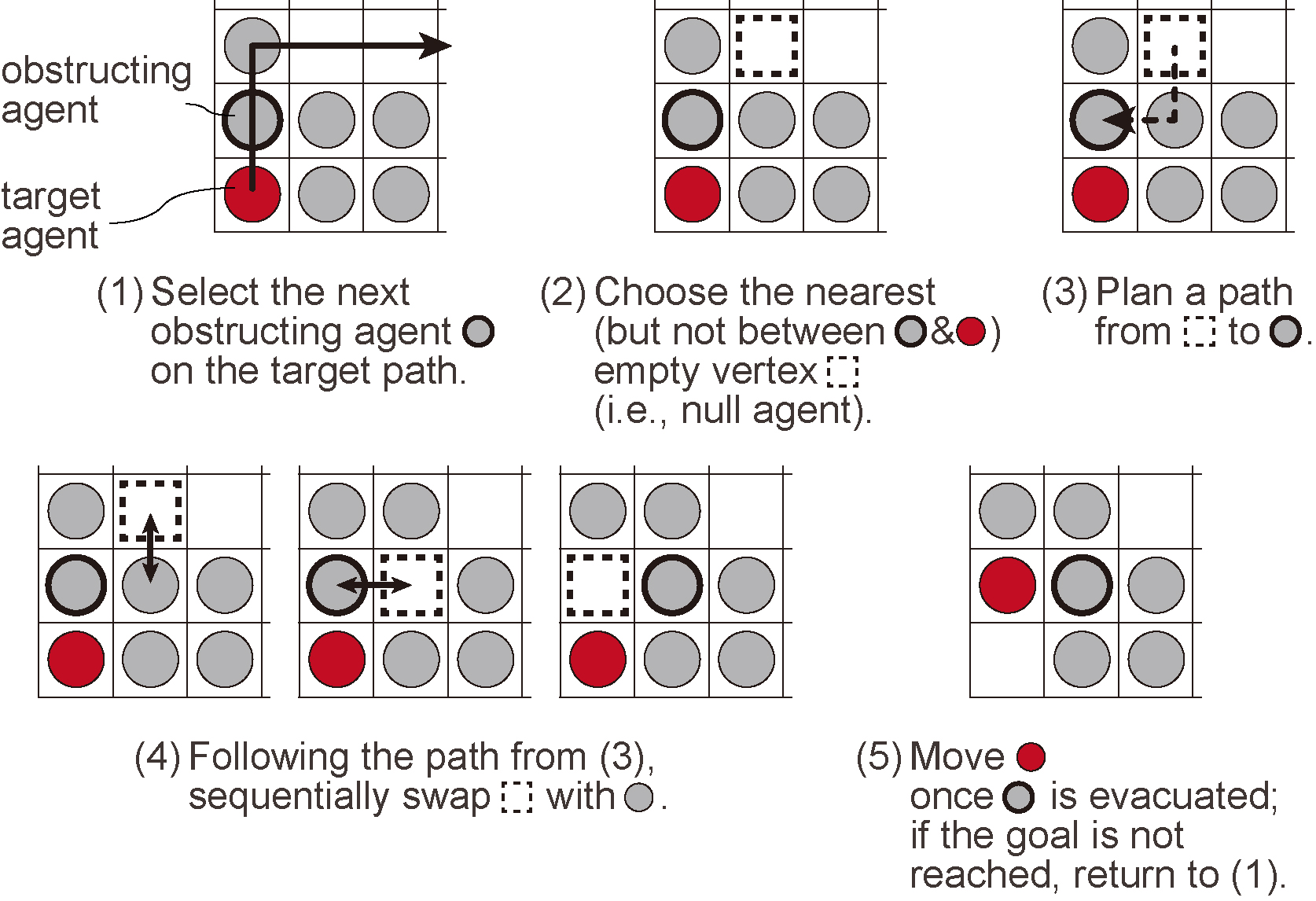}
  \caption{Second stage of PHANS: Null-agent swapping mechanism.}
  \label{null_agent_swapping}
\end{figure}

The obstructing agents identified on the target path in the first stage are evacuated sequentially.
This comprises five steps, as outlined in Fig.~\ref{null_agent_swapping}:
\begin{enumerate}
\item Select the next obstructing agent that blocks the movement of the target agent.
\item Choose the nearest (but not between the target and obstructing agents) empty vertex, termed the null agent.
\item Plan a path from the null vertex to the obstructing agent (null-agent path), ignoring potential collisions with the other obstructing agents.
\item Follow the null-agent path and swap the null agent with the neighboring agent (\textit{null-agent swapping}).
\item Move the target agent once the obstructing agent is evacuated; if the goal is not reached, repeat the process from Step 1.
\end{enumerate}
By iterating these steps, the obstructing agents are sequentially evacuated from the target path, enabling the target agent to progress toward its goal.

\subsection{Multiple Target Agents}
For multiple target agents, the first stage can be executed simultaneously for all targets.
Additionally, the second stage allows the simultaneous evacuation of multiple obstructing agents to enhance efficiency.
The pseudocode for this generalized approach is provided in Algorithm \ref{algorithm_tpd}.
We perform sequential A* path planning using prioritized planning (PP), giving higher priority to targets with longer start-goal distances computed via \eqref{eq:manhattan_distance}.

Lines 1--4 in Algorithm~\ref{algorithm_tpd} compute the distance from the start to the goal for each target agent in $A^\mathrm{tgt}$ using \eqref{eq:manhattan_distance} and then prioritize the planning of target paths with the longest distances.
As resolving conflicts during simultaneous path planning increases computational costs, paths are planned sequentially using PP. 

Lines 6--10 identify obstructing agents on the target paths and add them to set $B$ for subsequent processing.
Line 11 orders the obstructing agents in $B$ according to their remaining distance along the blocked target's path to its goal.
Agents that block a longer remaining segment are handled first to reduce the makespan.

Lines 12--16 assign a null agent to each obstructing agent and plan the path between them.
The null-agent assignment proceeds in order from the head of $B$, stopping if no null agents remain.
In general, the nearest null agent is assigned to each obstructing agent. 
However, because target agents are prioritized over obstructing agents, any null agents lying on the target agent's path between the target agent and the obstructing agent are reserved for the target agent.
Therefore, these reserved null agents are not assigned to obstructing agents.
Lines 18--21 move the target agent after the obstructing agent is evacuated.
If there are no collisions, lines 22--25 execute the sequential relocation of the obstructing agents along the paths planned with their assigned null agents.

\subsection{How Fast Is PHANS?}
The A* algorithm's worst-case time complexity is $\mathcal{O}\left(|E|+|V|\log|V|\right)$~\cite{fredman1987}.
The proposed method uses A* to compute the target path and repeatedly replans the paths for obstructing agents along that path.
With $m$ target agents and the longest target path of length $d$, the worst-case time complexity of the proposed method is $\mathcal{O}\left(m^2d^2(|E|+|V|\log|V|)\right)$, which accounts for repeated path recalculations.

Conversely, the methods proposed by Okoso et al.~\cite{okoso2022} and Makino et al.~\cite{makino2024} extend the graph $G=(V,E)$ along the time axis and solve the minimum-cost flow problem through edges for path planning.
With a time-axis size of $T$, their worst-case time complexity is $\mathcal{O}\left(2^{T|E|}\right)$.

Therefore, the proposed method operates in polynomial time, whereas the traditional approaches exhibit exponential time complexity.
Consequently, the computational cost of the proposed method is expected to be significantly lower than that of the conventional methods.

\begin{figure}[!t]
  \begin{algorithm}[H]
    \caption{PHANS}
    \label{algorithm_tpd}
      \begin{algorithmic}[1]
        \footnotesize
        \DefineClass Agent
        
        \textit{\textbf{pos}}: Tuple

        \textit{\textbf{goal}}: Tuple

        \textit{\textbf{dst}}: Integer

        \textit{\textbf{path}}: Sequence of Tuple
        \Require{
        \Statex $A^\mathrm{tgt}$ = \{(\textit{\textbf{pos}}: start position, \textit{\textbf{goal}}: goal position, \textit{\textbf{dst}}: None, \textit{\textbf{path}}: None) $\mid$ target agent $\in$ all target agents\} 
        \Statex $A^\mathrm{obs}$ = \{(\textit{\textbf{pos}}: start position, \textit{\textbf{goal}}: None, \textit{\textbf{dst}}: None, \textit{\textbf{path}}: None) $\mid$ obstructing agent $\in$ all obstructing agents\}\nonumber}
        \Ensure{Agents' movements}
        \State Sort $A^\mathrm{tgt}$ in descending order of distance between start and goal
        \ForAll {tgt in $A^\mathrm{tgt}$}
          \State tgt.path $=$ path from tgt.pos to tgt.goal
          \State tgt.dst  $= |\mathrm{tgt.path}|$ 
        \EndFor
        \While{not all target agents have reached their goals}
          \State $N$ = all null agents
          \State $B = \{\}$
          \ForAll {obs in $A^\mathrm{obs}$}
            \If {obs is on the target paths}
              \State $B$.add(\{\textit{\textbf{pos}}: obs.pos, \textit{\textbf{goal}}: None, \textit{\textbf{dst}}: len of path from obs.pos to its goal, \textit{\textbf{path}}: None\})
            \EndIf
          \EndFor  
          \State Sort $B$ in descending order of dst
          \ForAll {bagt in $B$}
            \If {one or more null agents in $N$}
              \State bagt.goal $=$ nearest (but not between the target and obstructing agent) null agent from bagt.pos
              \State bagt.path $=$ null-agent path from bagt.goal to bagt.pos
              \State remove the selected null agent from $N$
            \EndIf
          \EndFor 
          \While{no target agents have moved}
            \ForAll {tgt in $A^\mathrm{tgt}$}
              \If{$|\mathrm{tgt.path}|> 1$ and tgt.path[1] is null}
                \State move the tgt from tgt.path[0] to tgt.path[1]
                \State tgt.path $=$ tgt.path[1:] 
              \EndIf
            \EndFor
            \ForAll {bagt in $B$}
              \If{bagt.path[1] is null}
                \State move the bagt from bagt.path[0] to bagt.path[1]
                \State bagt.path $=$ bagt.path[1:]
              \EndIf
            \EndFor
          \EndWhile
        \EndWhile
      \end{algorithmic}
  \end{algorithm}
\end{figure}

\subsection{Why Does PHANS Terminate Within a Finite Number of Steps?}
Each target agent $a^\mathrm{tgt}_i$ has a target path of finite length, and the set of all obstructing agents on these paths is denoted as $B$.

For any vertex $v$, function $d_\mathrm{null}^t(v)$ returns the path length to the nearest empty vertex at time step $t$, where an empty vertex indicates no agent occupancy.
If $d_\mathrm{null}^t(v)=0$, then vertex $v$ is empty.

The proposed method prioritizes path planning for obstructing agents on the target paths by assigning a higher priority to those farthest from the goal.
By definition of the environment, at least one empty vertex exists; therefore, this vertex can be assigned to the obstructing agent with the longest path to its goal.
Let $v^\ast$ denote the vertex of the obstructing agent.
\begin{align}
    v^\ast = \underset{v\in B} {\operatorname{argmax}} \, dst(v, \mathrm{\,goal\ vertex\ of\ } v\mathrm{\text{'}s\ obstructing\ target}).
\end{align}

A finite-length path exists between $v^\ast$ and the nearest empty vertex (null agent), and null-agent swapping is performed at each time step along this path.
Consequently, $d^t_\mathrm{null}(v^\ast)$ decreases monotonically at each time step, i.e., $d_\mathrm{null}^t(v^\ast) > d_\mathrm{null}^{t+1}(v^\ast)$.
This process continues until $d^t_\mathrm{null}(v^\ast)$ reaches zero after a finite number of steps.
At this point, $v^\ast$ becomes empty, allowing the farthest target agent to advance to a newly vacated vertex.

Given the finite lengths of the target paths, all target agents reach their goals after a finite sequence of null-agent swaps and movements toward empty vertices.
Therefore, PHANS terminates within a finite number of steps.

\section{Numerical Experiments}
This section describes the four experiments conducted to evaluate the performance of the proposed method.
Experiments~1 and 2 compared PHANS with PIBT, EECBS, and ILP in small-scale simulation environments where these methods could be applied. 
Experiment~3 assessed the scalability of PHANS in large environments, demonstrating its effectiveness.
Additionally, Experiment~4 assessed the applicability of PHANS in complex environments with obstacles such as pillars. 
All experiments were conducted on an Intel Core i9-12900K processor with 128 GiB of RAM and Ubuntu 22.04.

\subsection{Settings for Comparison Methods}
We compared PHANS against ILP, PIBT, and EECBS. 
To the best of our knowledge, ILP is the only solver that optimizes the relocation of obstructing agents~\cite{okoso2022, makino2024}.
PIBT is a state-of-the-art solver among the fast suboptimal algorithms.
EECBS is an extension of the classical CBS. 
Li and Ma~\cite{li2023} used EECBS to relocate shelves in dense warehouse scenarios.
Setting the suboptimality factor $\omega=1.0$ ensures that EECBS operates in its optimal mode, thereby enabling a direct comparison of optimal solutions while simplifying the experimental setup.
Note that PIBT is far more scalable than EECBS~\cite{okumura2022}.

Unlike PHANS and ILP, PIBT and EECBS require predefined goals for agents whose paths are planned. 
In the experiments, we assigned the obstructing agents goals that return them to their starting positions. 
However, the obstructing agents stopped moving once all target agents reached their goals. 
In other words, the makespan was computed only for the target agents.

ILP defines target and obstructing agents separately, and paths for all agents are planned simultaneously. 
As in~\cite{okoso2022, makino2024}, we used Gurobi as the ILP solver.

PIBT can assign priorities to agents. 
Accordingly, we set each target’s priority equal to its start-goal distance and assigned a priority of $0$ to obstructing agents.
Because vanilla PIBT does not consider following conflicts, we extended it using the method described in~\cite{okumura2021}. 

\begin{figure}[!t]
  \centering
  \includegraphics[scale=0.12]{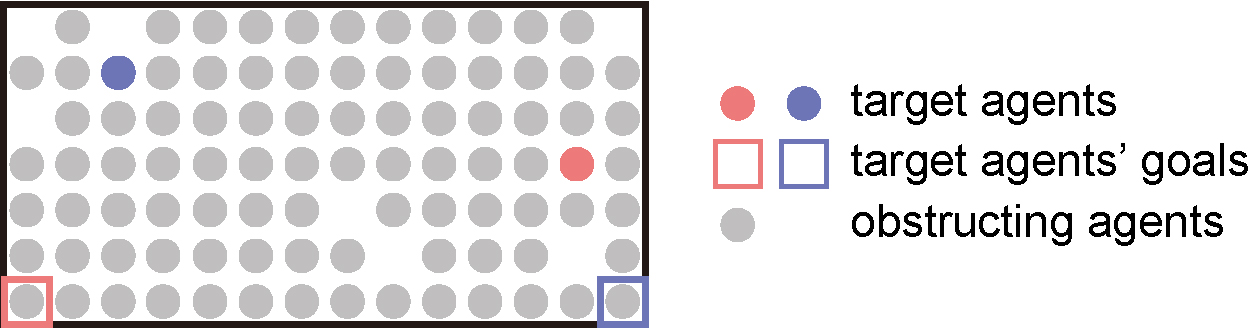}
  \caption{Simulation environment for Experiment~1. The grid size is $14\times 7$. Obstructing agents occupy $90\%$ of the cells.}
  \label{exp1_env}
\end{figure}

\begin{figure}[!t]
  \centering
  \includegraphics[width=\linewidth]{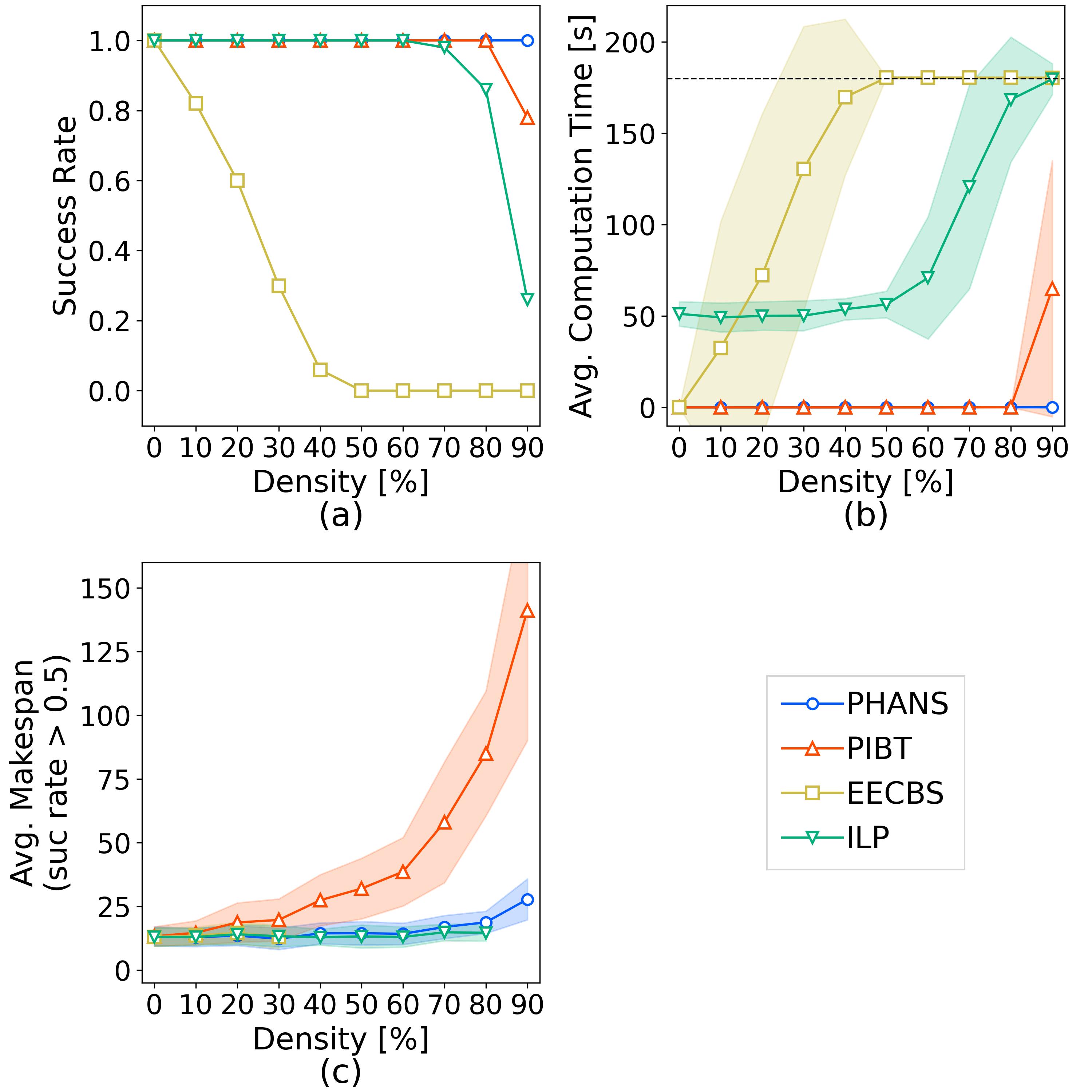}
  \caption{Comparison of (a) success rates, (b) computation time, and (c) makespans between PHANS and conventional methods in Experiment~1. The density is the proportion of cells occupied by obstructing agents.}
  \label{exp1_result}
\end{figure}

\subsection{Experiment 1}
We compared PHANS with conventional methods (PIBT, EECBS, and ILP) based on the success rate, computation time, and makespan.
The computation time was the time each method took to complete the path-planning process, with a maximum of $180$~s in each trial.

The simulation environment is shown in Fig.~\ref{exp1_env}.
The simulations involved two target agents, and the number of obstructing agents varied from $0\%$ to $90\%$ of the total number of cells.
The objective was to minimize the makespan and move the target agents to their destinations without collisions.

Fig.~\ref{exp1_result} shows the results of fifty experimental runs for each density, with different initial positions for the agents in each run.
The horizontal axis represents the density of obstructing agents.
Fig.~\ref{exp1_result}(a) shows the success rates, and Fig.~\ref{exp1_result}(b) presents the computation time.
Fig.~\ref{exp1_result}(c) presents makespan comparisons for scenarios where feasible solutions were identified for at least half of the trials.

\subsubsection{Success Rate and Computation Time}
Figs.~\ref{exp1_result}(a) and (b) show the success rate and computation time, respectively.
PHANS completed all path-planning tasks with the lowest computation times across various densities.
Even at a density of $90\%$, the average computation time remained low at $0.04$~s.

By contrast, ILP required an average computation time of approximately $50$~s, even in low-density environments.
This can be attributed to the time needed for preprocessing and constructing the time-expanded network and the inherently high computational cost of ILP.
Despite its high computational cost, ILP achieved high success rates, with values of $1.0$ up to a density of $60\%$ and $0.86$ at $80\%$, surpassing those of EECBS.

In low-density environments, the computation time of EECBS was less than that of ILP.
However, starting at $20\%$ density, the computation time of EECBS increased drastically, accompanied by a decline in the success rate.
EECBS was efficient in low-density environments. By contrast, high-density environments increased the collision resolution cost, leading to lower success rates. 

Compared with ILP and EECBS, PIBT required less computation time; at a density of $80\%$, it took only $0.11$~s.
The computation time increased considerably at a density of $90\%$, due to the computational complexity of PIBT.
PIBT does not know where empty cells exist prior to the evacuation of obstructing agents.
Agents in PIBT only know their goals, and the evacuating position (selecting empty cells) is not optimized.
To avoid following conflicts, PIBT performs a random search to determine the evacuating position.
If the probability that a cell is occupied is $p$, the probability of finding an empty cell in a given number of search attempts is $p^{\mathrm{times}-1}\cdot (1-p)$, where $p$ is assumed to be fixed and independent of previously visited paths.
At a density of $90\%$, the obstructing agents struggle to find the $10\%$ unoccupied cells.
The computational complexity at $90\%$ density is much higher than at lower densities, which consequently increases the computation time.

\subsubsection{Makespan}
Fig.~\ref{exp1_result}(c) compares the makespans of PHANS and conventional methods for scenarios where feasible solutions were identified for at least half of the trials.

PIBT has a much larger makespan than PHANS.
PIBT can resolve local conflicts (around the target agents) via backtracking.
However, backtracking cannot be applied to future conflicts, and moving the obstructing agents in advance is challenging.
Therefore, the number of local-conflict resolutions increases, which increases the makespan.

Fig.~\ref{exp1_result}(c) shows the makespan of ILP to be consistently smaller than that of PHANS.
This difference is particularly notable in high-density environments, where ILP's makespan was approximately $5$--$10\%$ shorter.
ILP is an optimal solver, whereas PHANS is suboptimal.
The target paths, selected null agents, and paths of the null agents may not be optimal in PHANS; consequently, PHANS has a larger makespan.

\subsection{Experiment 2}
\begin{figure}[!t]
  \centering
  \includegraphics[scale=0.12]{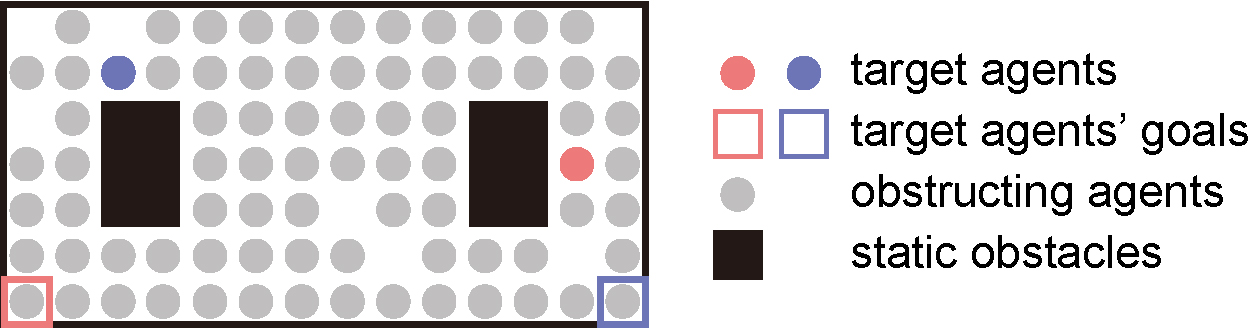}
  \caption{Simulation environment for Experiment~2. The grid size is $14\times 7$. Obstructing agents and static obstacles occupy $90\%$ of the cells.}
  \label{exp2_env}
\end{figure}

Experiment~1 was conducted in an environment without obstacles.
Experiment~2 was conducted in an environment with pillar-like obstacles (Fig.~\ref{exp2_env}).
We set the number of target agents to $2$ and the cell occupation density varied from $20\%$ to $90\%$ of the cells.
The densities included obstructing agents (varied) and static obstacles (fixed at $90$ cells).
As in Experiment~1, the maximum computation time for each trial was set to $180$~s.

\subsubsection{Success Rate and Computation Time}
The success rates and computation times (Figs.~\ref{exp2_result}(a) and (b)) of PHANS, EECBS, and ILP were similar to those in Experiment~1.
PIBT achieved shorter computation times than in Experiment~1, particularly at $90\%$ density.
PIBT moved obstructing agents to empty cells via backtracking. 
As static obstacle cells were excluded from the search, the search space was reduced compared with environments without obstacles, ultimately resulting in a smaller computation time and improved success rate.

\subsubsection{Makespan}
\begin{figure}[!t]
  \centering
  \includegraphics[width=\linewidth]{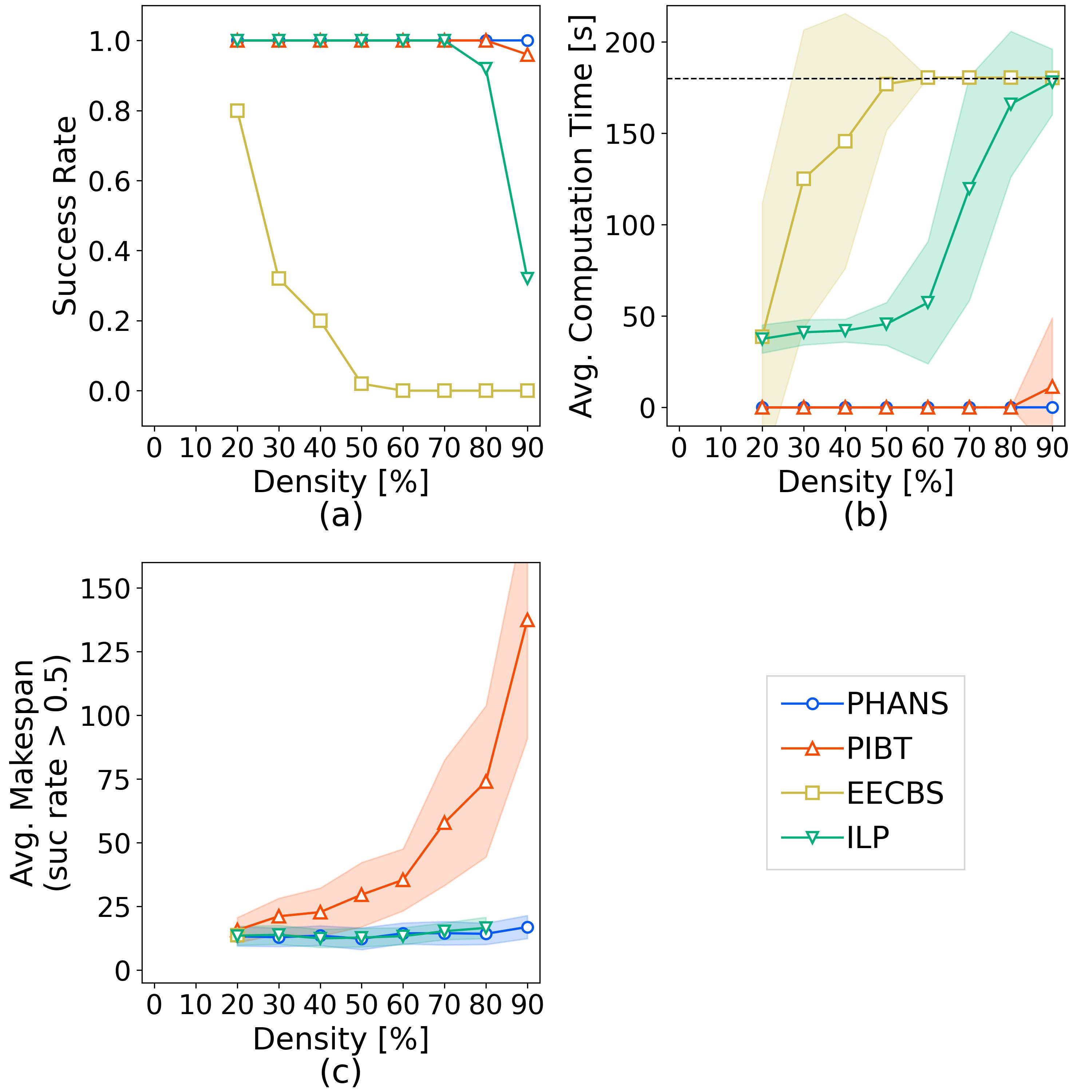}
  \caption{Comparison of (a) success rates, (b) computation time, and (c) makespans between PHANS and conventional methods in Experiment~2. The density is the proportion of cells occupied by obstructing agents and static obstacles.}
  \label{exp2_result}
\end{figure}

The makespans (Fig.~\ref{exp2_result}(c)) of EECBS and ILP were similar to those in Experiment~1, whereas PHANS and PIBT achieved shorter makespans under certain conditions.
This is presumably because the presence of static obstacles reduces the search space, making it easier to evacuate obstructing agents to favorable locations.

\begin{figure}[!t]
  \centering
  \includegraphics[scale=0.12]{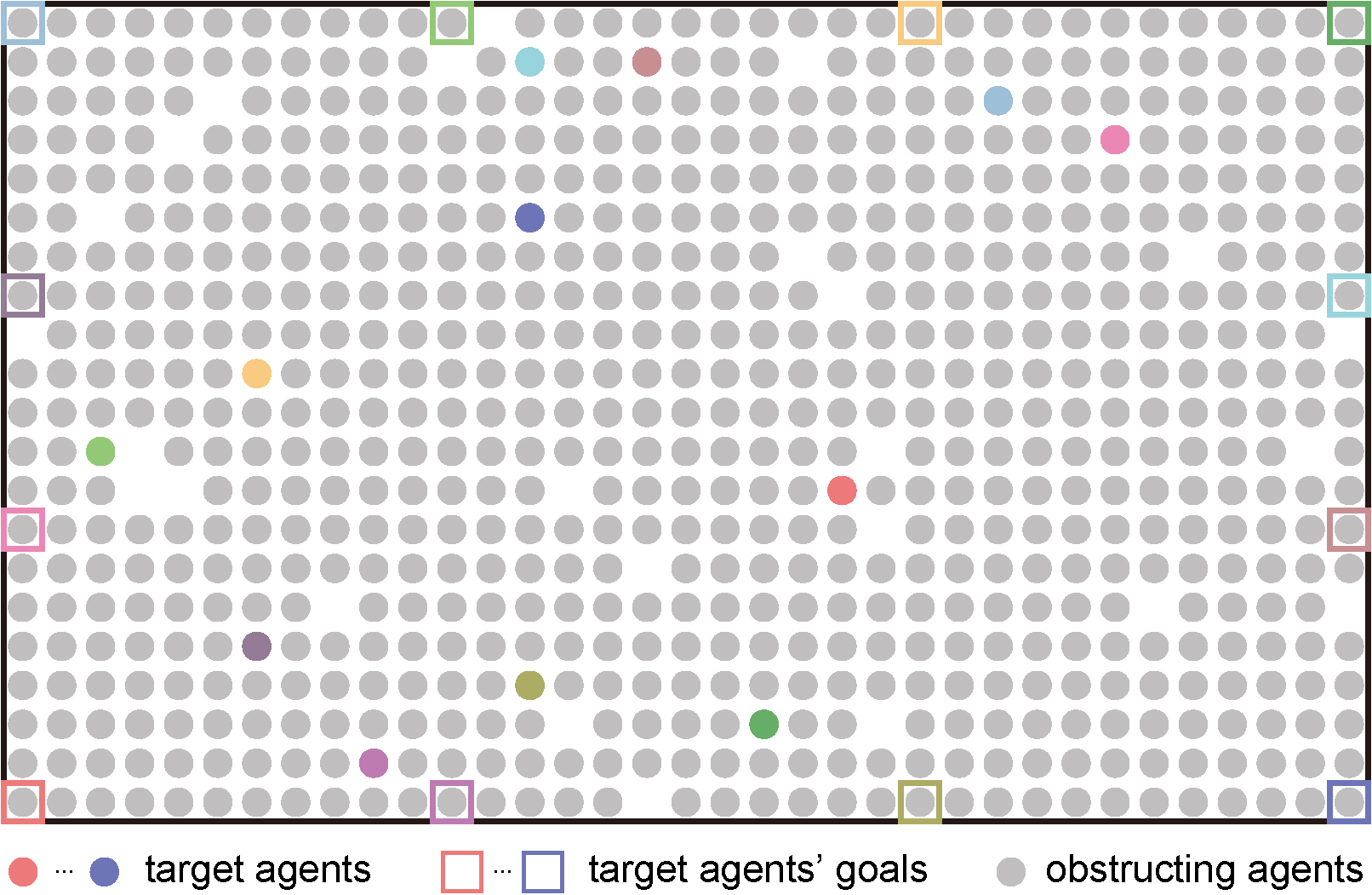}
  \caption{Simulation environment for Experiment~3. The grid size is $35\times 21$. Obstructing agents occupy $95\%$ of the cells.}
  \label{exp3_env}
\end{figure}

\begin{figure}[!t]
  \centering
  \includegraphics[width=\linewidth]{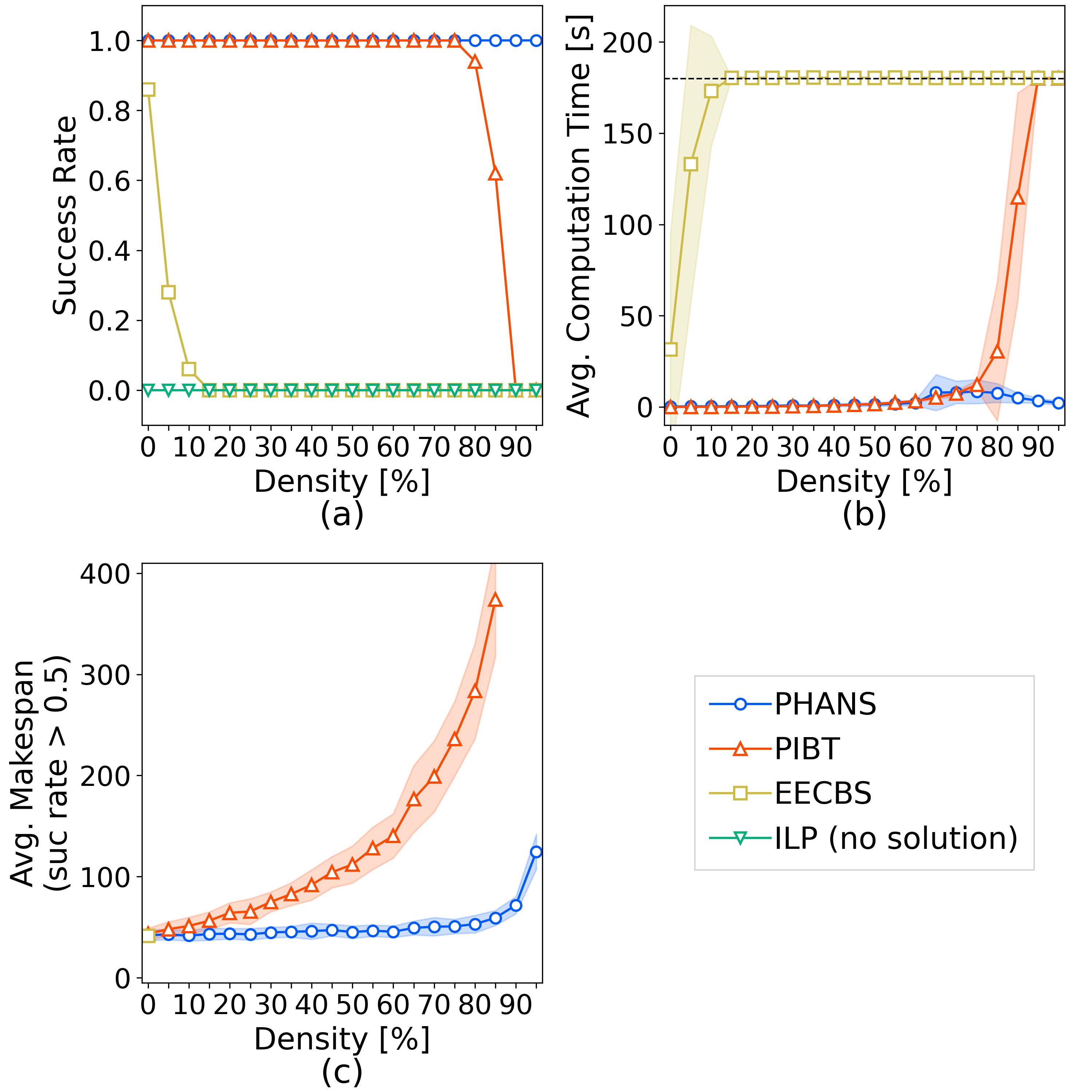}
  \caption{Comparison of (a) success rates, (b) computation time, and (c) makespans between PHANS and conventional methods in Experiment~3. The density is the proportion of cells occupied by obstructing agents.}
  \label{exp3_result}
\end{figure}

\subsection{Experiment 3}
We evaluated the scalability of PHANS by testing its performance in an environment larger than that considered in Experiments~1 and 2.
Specifically, we assessed the makespan and computation time across varying densities of obstructing agents.

Fig.~\ref{exp3_env} shows the simulation environment, which is a free-location automated warehouse in which the picking locations were designed as the goals for the target agents along the perimeter. 
We set the number of target agents to 12, and the number of obstructing agents varied from occupying $0\%$ to $95\%$ of the cells. 
Fifty trials were performed at each density. 
As in Experiments~1 and 2, the maximum computation time for each trial was set to $180$~s for all methods. 
Additionally, for the ILP baseline only, we ran a separate feasibility check at $95\%$ density with an extended limit of $3600$~s; ILP failed to find a solution within this time. 
All results reported in this section use the $180$~s cap.

\begin{figure}[!t]
  \centering
  \includegraphics[scale=0.12]{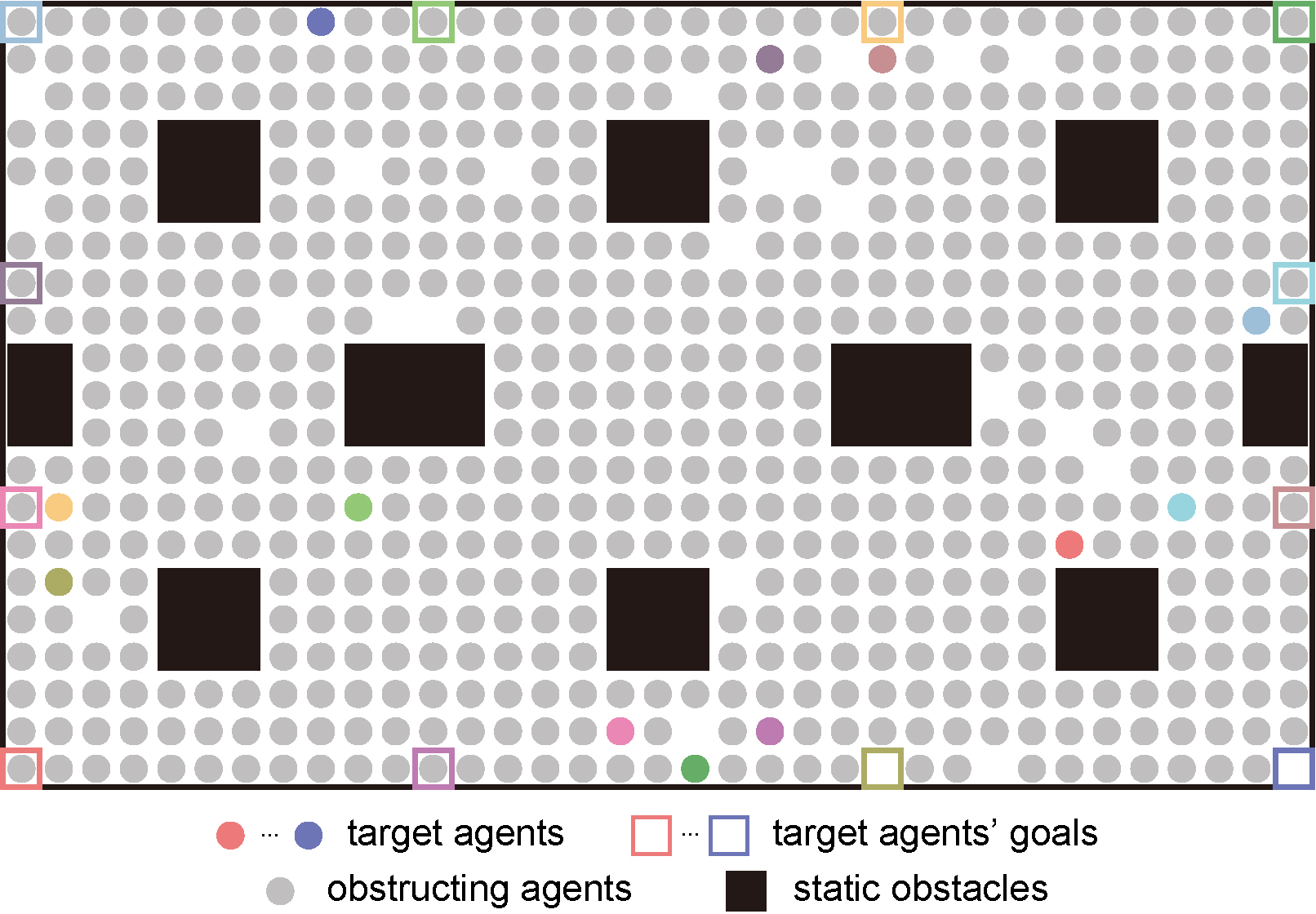}
  \caption{Simulation environment for Experiment~4. The grid size is $35\times 21$. Obstructing agents and static obstacles occupy $95\%$ of the cells.}
  \label{exp4_env}
\end{figure}

\begin{figure}[!t]
  \centering
  \includegraphics[width=\linewidth]{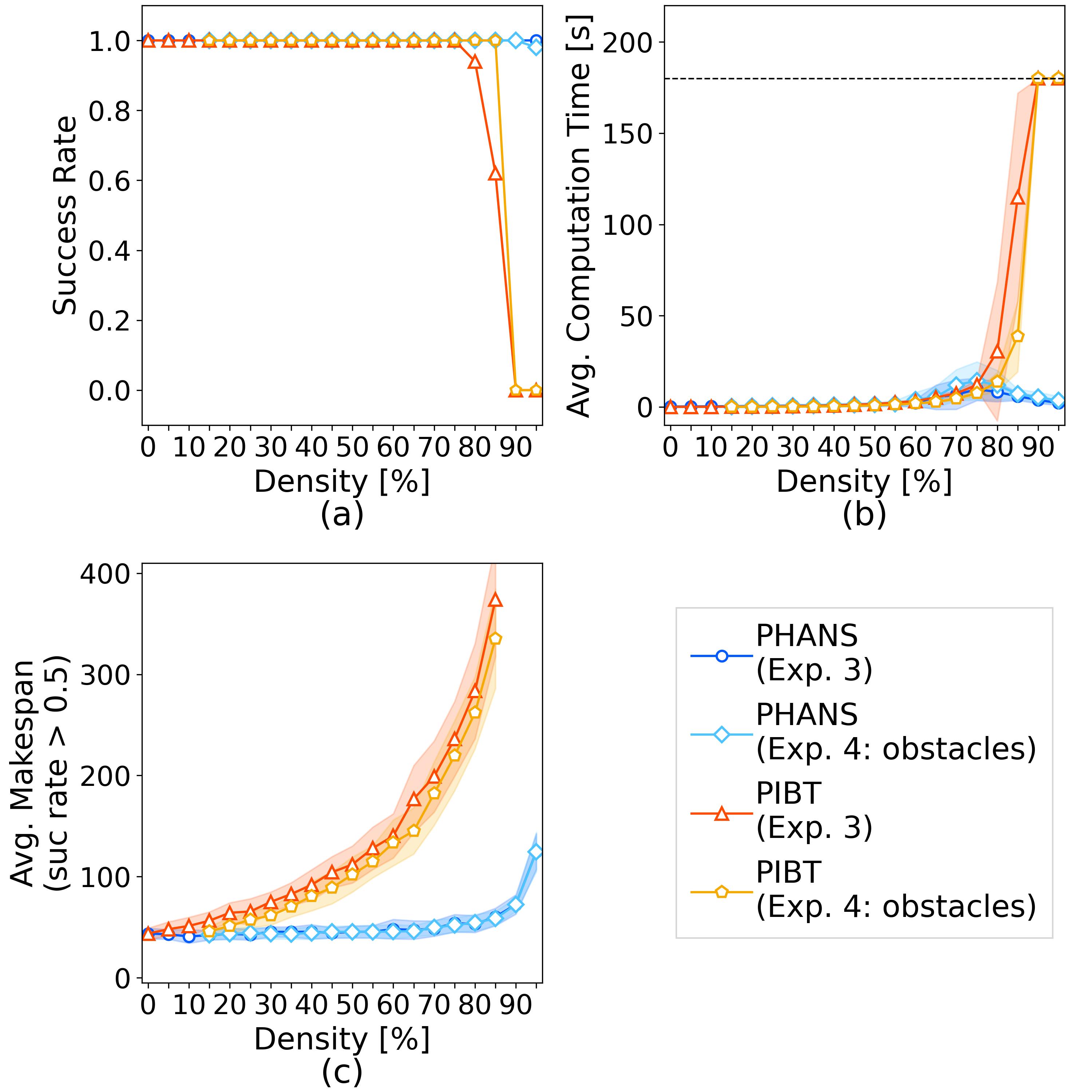}
  \caption{Comparison of (a) success rates, (b) computation time, and (c) makespans between PHANS and PIBT in Experiment~4. The density is the proportion of cells occupied by obstructing agents and static obstacles.}
  \label{exp4_result}
\end{figure}

\subsubsection{Success Rate and Computation Time}
Figs.~\ref{exp3_result}(a) and (b) show the success rate and computation time, respectively.
PHANS completed all path-planning tasks with minimal computation time across various densities.
Similar to Experiment~1, even at the highest density ($95\%$), the average computation time remained low at $2.07$~s.

Fig.~\ref{exp3_result}(b) indicates that higher obstructing agent densities generally led to longer computation times.
Increasing the number of obstructing agents on the target paths requires more evacuations.
This increases the number of calculations required for null-agent paths, which increases the computation time.

Fig.~\ref{exp3_result}(b) shows that the computation time of PHANS decreased when the density of obstructing agents exceeded $80\%$.
Even at a density of $95\%$, the average computation time was only $2.07$~s.
PHANS assigns priorities to the obstructing agents, and null agents are systematically allocated until no further assignments are possible (Algorithm \ref{algorithm_tpd}, lines 6--16).
At extremely high densities, the limited availability of null agents simplifies the assignment process and reduces the computation time.
By contrast, ILP faces tighter global coupling on the time-expanded network; EECBS resolves more conflicts as density rises; and PIBT incurs heavier backtracking when feasible local moves vanish. 
Thus, only PHANS exploits the lack of empty cells to shrink the assignment space, which lowers the computation time at high occupation density.

The success rates and computation times of PIBT, EECBS, and ILP were similar to those in Experiment~1.
The computational cost of PIBT increases because, in extremely high-density environments, the probability of having an empty neighboring cell decreases, which increases the backtracking computational cost.  
In ILP, the costs of preprocessing and constructing the time-expanded network increase exponentially with the scale of the environment. 
Consequently, even when the computation time limit was relaxed from 180~s to 3600~s, ILP failed to find a feasible solution.

\subsubsection{Makespan}
Fig.~\ref{exp3_result}(c) compares the makespans of PHANS and conventional methods.

As in Experiment~1, the makespan of PHANS was significantly smaller than that of PIBT.
In large-scale environments, target paths become longer, and the number of obstructing agents that need to be moved increases.
At this point, moving the obstructing agents in advance has a significant effect.

The PHANS makespan remained below $60$ for densities less than $90\%$. However, it increased drastically at higher densities.
This increase is attributed to the additional time required to evacuate more obstructing agents, which leads to longer waiting times for the target agents and thus longer makespans.

\subsection{Experiment 4}
We evaluated the applicability of PHANS in complex environments with obstacles such as pillars, which are common in large warehouse environments. 
For Experiment~4 (Fig.~\ref{exp4_env}), we set static obstacles as in Experiment~2 and assessed the makespan and computation time across varying densities.
We set the number of target agents to 12, whereas the occupation density varied from $15\%$ to $95\%$ of the cells.
The densities included obstructing agents (varied) and static obstacles (fixed at $90$ cells).
Fifty trials were performed for each density. 
As in Experiments~1--3, the maximum computation time for each trial was $180$~s.

\subsubsection{Success Rate and Computation Time}
As in Experiment~3, PHANS demonstrated high computational efficiency in the obstructed environment considered in Experiment~4.
Compared with PIBT, PHANS successfully planned paths in most trials with significantly shorter computation time (Figs.~\ref{exp4_result}(a) and (b)).
However, at a density of $95\%$, PHANS failed to plan a path in one trial. 
PHANS prioritizes target agents based on their distance from the goal and statically determines target paths.  
Owing to high-priority target agents and static obstacles, a deadlock occurred in one trial during the evacuation of obstructing agents, causing path-planning failure.

\subsubsection{Makespan}
In the obstructed environment of Experiment~4, the PHANS makespans were similar to those in Experiment~3; however, the computation time increased at densities of $70$--$80\%$ (Fig.~\ref{exp4_result}(b) and (c)). 
The presence of static obstacles increases the path-finding complexity for evacuating obstructing agents, increases the computational costs, and extends the paths for evacuating obstructing agents.  
However, by moving obstructing agents in advance based on the movement of target agents, the makespan was not significantly affected.

Interestingly, the makespan and computation time for PIBT were smaller in Experiment~4.
As in Experiment~2, this is likely because the presence of static obstacles reduces the search space explored by PIBT during backtracking.

\section{Conclusion}
This study defines MAPF-HD and introduces a new heuristic method, PHANS, to solve MAPF-HD efficiently. 
PHANS operates in two phases: first, it plans paths for the target agents, and then it sequentially evacuates obstructing agents to clear the path. 
This approach involves swapping positions between agents and empty vertices, which we refer to as null agents. 

Numerical experiments demonstrated that existing methods (e.g., ILP, EECBS, and PIBT) often incur substantial computational overhead or yield suboptimal makespans as the density of obstructing agents increases. 
By contrast, PHANS consistently delivers fast computation times and maintains competitive makespans even under extreme density conditions and in environments with static obstacles.

However, PHANS has limitations in certain complex scenarios. 
For instance, its performance can degrade in environments with a large number of static obstacles or a high ratio of target agents; nonetheless, such conditions are less typical in automated warehouses and valet parking applications.
PHANS does not explicitly handle dynamic obstacles; however, its rapid planning speed allows for frequent re-planning when such obstacles appear.

Overall, our results highlight PHANS as a scalable and practical approach for real-world applications such as high-density warehouse logistics and valet parking systems. 
Future research will focus on adaptive target-path planning and null-agent assignment strategies.
Additionally, we will extend PHANS to handle more complex scenarios and arbitrary graphs.


\end{document}